\documentclass[preprint,showpacs,preprintnumbers,amsmath,amssymb,showkeys,aps]{revtex4}
\usepackage{graphicx}
\usepackage{dcolumn}
\usepackage{bm}
\usepackage[ansinew]{inputenc}
\usepackage[T1]{fontenc}

\input{tcilatex}

\begin{document}
\preprint{}
\title{Manipulation of the Spontaneous Emission Dynamics of Quantum Dots
in 2D Photonic Crystals}
\author{A. Kress}
\email{kress@wsi.tum.de}
\author{F. Hofbauer}
\author{N. Reinelt}
\author{M. Kaniber}
\author{H. J. Krenner}
\author{R. Meyer}
\author{G. B\"ohm}
\author{J. J. Finley}

\affiliation{Walter Schottky Institut and Physik Department, Technische Universit\"at
M\"unchen, Am Coulombwall 3, D-85748 Garching, Germany}
\date{\today}

\begin{abstract}
We demonstrate the ability to control the spontaneous emission
dynamics of self-assembled quantum dots via the local density of
optical modes in 2D-photonic crystals.  We show that an incomplete
2D photonic bandgap is sufficient to significantly lengthen the
spontaneous emission lifetime ($>2\times$) over a wide bandwidth
($\Delta\lambda\geq40$ nm). For dots that are both
\textit{spectrally} and \textit{spatially} coupled to strongly
localized ($V_{mode}\sim1.5(\lambda/n)^{3}$), high $Q\sim2700$
optical modes, we have directly measured a strong Purcell enhanced
shortening of the emission lifetime $\geq5.6\times$, limited only
by our temporal resolution.  Analysis of the spectral dependence
of the recombination dynamics shows a maximum lifetime shortening
of $19\pm4$.  From the directly measured enhancement and
suppression we show that the single mode coupling efficiency for
quantum dots in such structures is at least $\beta=92\%$ and is
estimated to be as large as $\sim97\%$.
\end{abstract}

\maketitle

\preprint{APS/123-QED}


Designer photonic materials fabricated from periodic dielectrics
provide a direct route toward achieving complete control of the
spontaneous emission of solids.\cite{YablonovichJohn87,Lodahl04}
The ability to manipulate the strength of the light-matter
coupling in this way lies at the very heart of modern optics, with
a variety of potential applications ranging from integrated
photonics\cite{Krauss99} to fundamental quantum
optics.\cite{Knill01,Lutkenhaus00,Yoshie04,REI04}  One of the most
remarkable quantum optical phenomena is the deterministic
generation of single photons from isolated quantum emitters.  In
this context, individual semiconductor quantum dots (QDs) are
ideal solid state emitters due to their high radiative efficiency,
stability and ease of incorporation into active
devices.\cite{Michler00,Solomon01,Pelton02,Hors03}   However, for
QDs embedded within an isotropic semiconductor the single photon
extraction efficiency is extremely low ($\eta_{ex}<1\%$) limiting
their realistic potential for applications in quantum information
science.\cite{Gisin02,Lutkenhaus00}  This problem can be addressed
by locating QDs within optical cavities and utilizing the Purcell
effect\cite{Purcell46} to funnel single photons into a single
optical mode for collection.\cite{Pelton02}  Using such
approaches, $\eta_{ex}$ larger than a few percent have been
reported for single dots incorporated into pillar
microcavities.\cite{Pelton02}  Cavities realized using photonic
crystals (PCs) may provide maximum flexibility to tune the local
density of photon states over a much wider bandwidth and achieve
full control of the spontaneous emission via the strength of the
local vacuum field fluctuations.\cite{Lodahl04}   Furthermore,
strongly localized modes in PCs combine a planar geometry with
high quality factors ($Q=\omega\tau_{photon}$) and small effective
mode volume ($V_{eff}$)\cite{Vuckovic03,Painter99,Akahane03},
potentially advantageous properties for achieving strong Purcell
enhancement and realizing \textit{efficient} QD based single
photon emitters.

In this paper we demonstrate control of the QD spontaneous
emission dynamics in such 2D PCs.  For dots that are both
\textit{spectrally} and \textit{spatially} coupled to high Q,
strongly localized cavity modes we directly measure a pronounced
shortening of the emission lifetime ($\geq5.6\times$), limited
only by the temporal resolution of our detection system.  Analysis
of the spectral dependence of the decay rate as a function of
emitter-cavity detuning shows that the maximum enhancement is as
large as $19\pm4$.  A strong ($>2\times$) reduction of the
emission decay time is observed over a wide bandwidth for dots
detuned from the cavity modes, demonstrating that a partial,
TE-bandgap is sufficient to tailor the QD spontaneous emission
dynamics.  From the directly measured enhancement and suppression
of the spontaneous emission lifetime we extract a single mode
coupling efficiency for dots in these PCs of $\beta=92\%$, the
spectrally dependent measurements indicating that it may become as
large as $\beta\sim97\%$.

The samples investigated consisted of a $d=400$ nm thick
Air-GaAs-Air slab waveguide into which a 2D photonic crystal (PC)
is defined by fabricating a triangular lattice of air holes in the
GaAs waveguide and defining a suspended membrane by removing an
underlying AlAs layer using wet chemical etching.  A single layer
of nominally In$_{0.5}$Ga$_{0.5}$As QDs was incorporated into the
center of the GaAs waveguide core as an internal light source. Low
mode volume microcavities were formed by introducing single
missing hole point defects in the hexagonal lattice of holes,
realizing $H1$ resonators. The PCs have a periodicity of $a=300$nm
and the air hole radius ($r$) was varied to tune the cavity mode
energies through the inhomogeneously broadened spectrum of QD
groundstates and control the width of the photonic bandgap.   From
a QD areal density of $\sim200\mu m^{-2}$ and the detection spot
size of $\sim1\mu m^{2}$ we estimate that only a small number of
dots ($\sim5$) are spectrally coupled to the high Q cavity modes.

\begin{figure}[tbhp]
    \begin{center}
       \includegraphics[width=1.0\columnwidth]{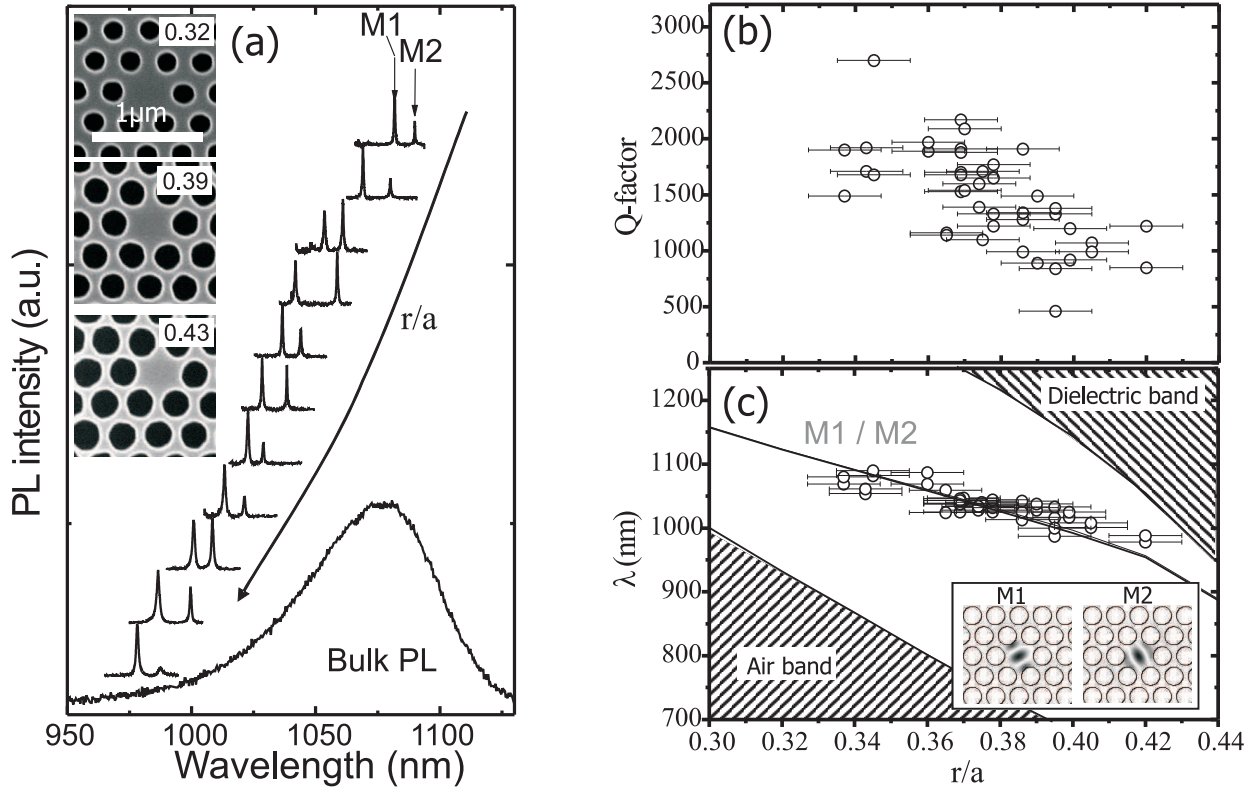}
   \end{center}
   \caption{\label{fig1} (a) PL spectra recorded from a series of $H1$
PC cavities as a function of the ratio of hole radius $r$ to
periodicity $a$.  (inset) The SEM images show typical cavities.
The ensemble PL is shown for comparison. (b) Measured Q factors
for the cavity modes of $>40$ cavities as a function of $r/a$. (c)
Calculations of the $TE-$polarized $3D$ bandstructure with $a=300$
nm showing the localized dipole like modes $M1$ and $M2$ inside
the photonic bandgap and the experimental data. (inset) Calculated
electric field profile of the cavity modes.}
\end{figure}

Spatially resolved optical measurements were performed at $T=10K$
using a confocal micro photoluminescence ($\mu PL$) system that
provides a spatial resolution $\sim 1\mu m$.  For $CW$
measurements, the samples were excited using a HeNe laser and the
resulting $PL$ signal was dispersed by a $0.55$m imaging
monochromator and detected using a nitrogen cooled Si-CCD camera.
Time resolved measurements were performed by exciting the sample
using $\sim50$ ps duration pulses at $\lambda =658$ nm and
detecting the temporal decay of the resulting luminescence using a
single photon Si-avalanche photodetector and time correlated
photon counting electronics.  The maximum temporal resolution
provided by this system is $\sim150ps$ after deconvolution with
the system response function, much shorter than the intrinsic
ground state radiative lifetime of our QDs ($\tau_{0}\sim0.8ns$).

Figure 1a compares an ensemble $PL$ spectrum with $\mu PL$ spectra
recorded from a series of $H1$ cavities as $r/a$ is increased
systematically from $0.33 - 0.42$.  Over this range of parameters,
the $\mu PL$ spectra reveal a prominent doublet, labeled $M1$ and
$M2$ in Fig. 1, corresponding to dipole like cavity modes
orientated along the $\Gamma-M$ and $\Gamma-K$ crystal directions
(see inset - Fig. 1c).  Under the present strong excitation
conditions ($P_{ex}\sim100Wcm^{-2}$) the PL intensity is
determined by the QD spontaneous emission lifetime and a
$50\times$ enhancement of the PL intensity is observed for dots
spectrally on resonance with the cavity modes when compared with
dots that are detuned.  This observation indicates the presence of
pronounced cavity $QED$ effects, an expectation confirmed by our
time resolved measurements presented below.

Fig. 1b shows the cavity mode Q-factors deduced for over 40
structures plotted as a function of $r/a$.  A significant increase
of Q from $\sim500$ to $\sim2700$ is observed as $r/a$ is reduced
from $0.42$ to $0.33$.  This can be explained by considering the
position of the cavity modes within the $TE$-photonic
bandgap.\cite{Painter99}   Fig. 1c shows the calculated
bandstructure for our structures as a function of $r/a$.\cite{MIT}
The continuum \textit{dielectric} and \textit{air} bands are
marked by the shaded regions, together with the $TE$ photonic
bandgap and the $M1$ - $M2$ doublet (solid
lines).\cite{PhotonicCrystalBook}  The calculated wavelength of
the cavity modes and its dependence on $r/a$ are in good
quantitative agreement with our measurements, confirming their
identification.\cite{modes}  The calculations presented in Fig. 1c
show that $M1$ and $M2$ shift progressively deeper into the
photonic bandgap as $r/a$ is reduced.  As a consequence, the modes
couple more weakly to the dielectric band continuum resulting in
the observed enhancement of the Q-factor.\cite{Painter99}  We now
shift our attention to the emission dynamics of QDs whose emission
frequency lie throughout the $TE$-photonic bandgap, both in and
out of resonance with the highest Q cavity modes.

The maximum photon lifetime in our cavities is
$\tau_{photon}=Q_{max}/\omega\sim 2$ ps, much shorter than the
typical QD spontaneous emission lifetime ($\tau_{0}\sim0.8$).
Furthermore, since the QD homogeneous linewidth is much narrower
than the cavity mode ($\Delta\lambda_{c}=\lambda_{c}/Q\sim0.5nm$,
c.f. $\Delta\lambda_{QD}\ll0.1nm$ \cite{Bayer02}) the light matter
coupling remains in the perturbative regime and can be described
by the Purcell effect.\cite{Purcell46,Gerard99} In this case, for
an ideal emitter on resonance with the cavity mode the spontaneous
decay lifetime is reduced by a factor
$F_{p}=3Q/(4\pi^{2}V_{mode})$, where $V_{mode}$ is the effective
volume of the cavity mode in units of $(\lambda_{c}/n)^{3}$.  For
the cavities discussed here, we calculate $F_{p}\sim100$, for
$Q\sim2000$ and $V_{mode}\sim1.5(\lambda/n)^{3}$, in good accord
with the $\sim50\times$ total enhancement of the emission
intensity observed for QDs spectrally in resonance with the cavity
modes (see Fig. 1a). However, to unambiguously separate the
influence on the QD radiative lifetime of the local density of
photonic states from simple improvements of the collection
efficiency due to the PC, we performed time resolved measurements.

\begin{figure}[tbhp]
\begin{center}
   \includegraphics[width=1.0\columnwidth]{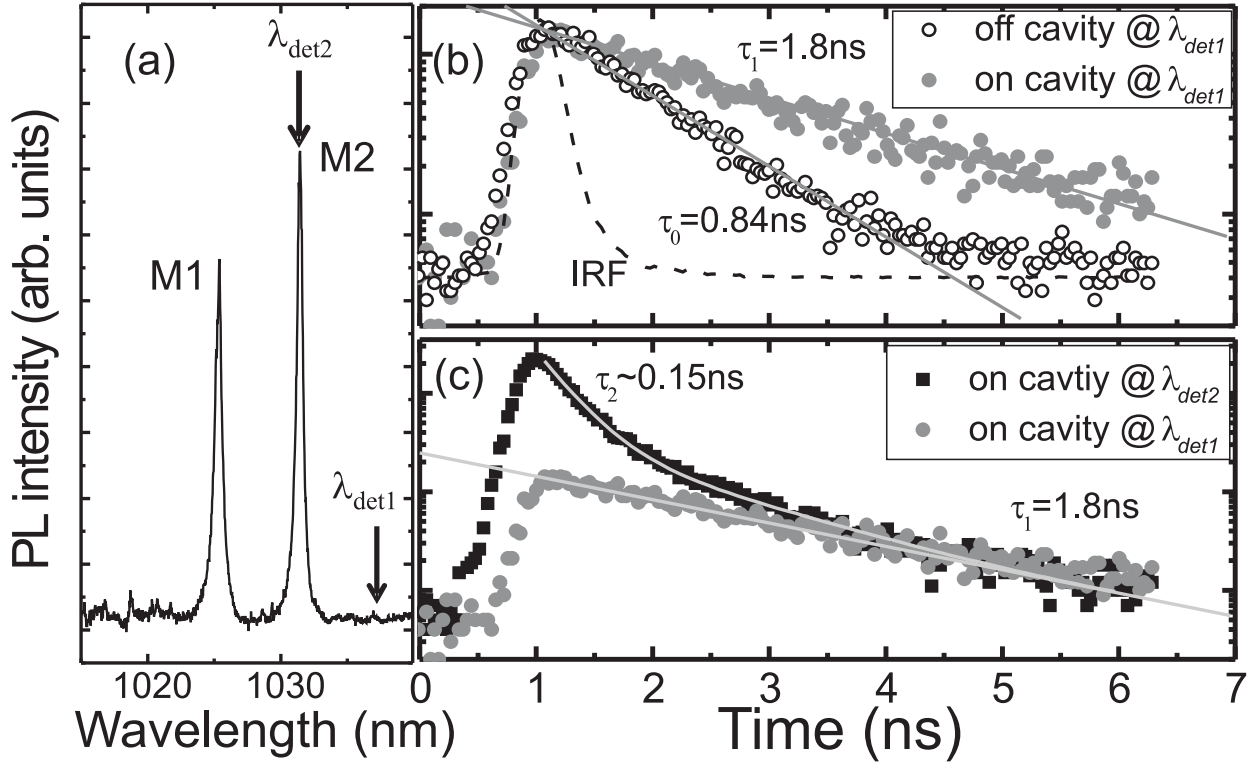}
\end{center}
   \caption{\label{fig2} (a) PL spectrum of the selected cavity.  (b) Comparison of decay
transients recorded from QDs detuned from the cavity modes
($\lambda_{det1}$=1037nm) either (i) within the unpatterned GaAs
membrane ($\tau_{0}$ - open circles) or (ii) from the PC
($\tau_{1}$ - filled circles).  The dashed line is the instrument
response function of our setup ($IRF$). (c) Decay transients
recorded from the PC $H1$ cavity both \textit{in resonance} with
$M2$ at $\lambda_{det2}$=1031.5 nm (filled squares) and detuned at
$\lambda_{det1}$=1037nm (filled circles).}
\end{figure}

A PL spectrum from the cavity selected for time resolved studies
is presented in Fig. 2a, showing cavity modes suitable for
detection using our silicon based detection system
($\lambda_{M1}=$1025.4 nm with $Q_{M1}=$1500 and $\lambda_{M2}=$
1031.5nm with $Q_{M2}=$1950).  We compared $\mu$PL decay
transients recorded both \textit{in} and \textit{out} of resonance
with the cavity modes with the \textit{intrinsic} QD dynamics
measured on the unpatterned GaAs membrane without the PC
($\tau_{0}$). Figure 2b compares raw time resolved data recorded
from QDs in the cavity, but strongly detuned from the cavity mode
(filled circles) with reference data recorded at the same
wavelength ($\lambda_{det1}=1037$ nm) from the unpatterned
membrane (open circles).  For both transients, we observe
monoexponential decays with time constants of $\tau_{1}=1.8\pm0.1$
ns and $\tau_{0}=0.84\pm0.05$ ns, respectively.  The QDs located
within the cavity have much longer decay times
($\tau_{1}/\tau_{0}\sim2$) when compared with dots in the pure
membrane, indicating the presence of a gap in the local photonic
density of states due to the 2D photonic bandgap.\cite{Lodahl04}
This suggestion is further substantiated by our spectrally
resolved measurements presented below.

Figure 2c compares a decay transient recorded from dots in the
cavity, here recorded in-resonance with $M2$ (filled squares),
with the data of Fig. 2b detuned strongly by $\sim5$nm to longer
wavelength (open circles).  In contrast with all dynamics
discussed until now, the decay transient recorded on resonance
with $M2$ shows a clear double exponential decay, as confirmed by
a fit of $I(t)=A\exp(-t/\tau_{2})+B\exp(-t/\tau_{1})$ shown on the
figure. The longer time constant ($\tau_{1}=1.8\pm0.1$ ns) is
identical to that discussed above for QDs spectrally detuned from
the cavity mode, whereas the faster transient
($\tau_{2}\sim0.15ns$) is limited by the time resolution of our
setup.  We identify this behavior as arising from a strong Purcell
enhanced shortening of the emission time, compared with
$\tau_{0}$, measured for dots that are both \textit{spectrally}
and \textit{spatially} on resonance with the cavity mode.  From
figure 2b and 2c we obtain already a factor
$\tau_{0}/\tau_{2}\geq5.6\pm0.3$, while the decay time $\tau_{2}$
is limited by the system time resolution (see instrument response
function in Fig. 2b). The longer decay transient $\tau_{1}$ is
identified as arising from QDs that are spectrally on resonance
with $M2$ but do not couple to the cavity mode due to their
position outside the cavity in the body of the PC. From these
directly measured decay times for coupled ($\tau_{2}$) and
uncoupled ($\tau_{1}$) dots (see figure 3) we obtain ratios of
$\tau_{1}/\tau_{2}=12\pm1$, defining a single mode coupling
efficiency $\beta=1-(\tau_{2}/\tau_{1})\sim92\%$ for dots placed
both spectrally and spatially on resonance with the cavity mode.
This figure of merit provides significant promise for the
realisation of efficient, QD based, single photon sources based on
PC cavities.\cite{Krauss99,Pelton02,Hors03}

\begin{figure}[tbhp]
    \begin{center}
       \includegraphics[width=0.9\columnwidth]{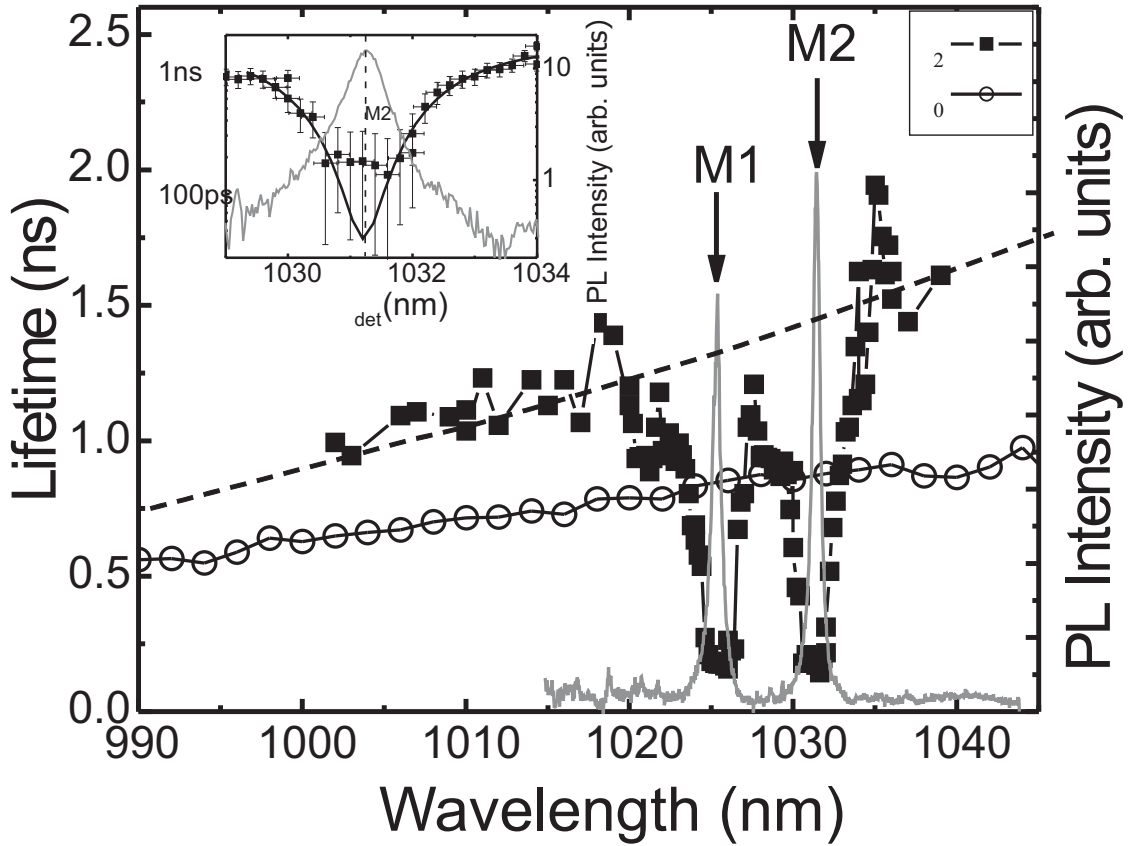}
   \end{center}
   \caption{\label{fig3}
   Spectral dependence of the decay lifetimes for dots within the PC (filled squares) and within
the unpatterned membrane (open circles).  The dotted line
represents a guide to the eye showing the stronger suppression of
the spontaneous emission for wavelengths closer to
$\lambda_{mid}^{TE}$. (inset) Fit of eqn 1 to the spectral
dependence of the decay lifetime.}
\end{figure}

Figure 3 compares the spectral dependence of the QD decay time in
the membrane, but away from the PC ($\tau_{0}$ - open circles) and
the dominant decay time for dots in the cavity (filled squares),
representing the faster of the two time constants extracted from
the biexponential fit. A reference PL spectrum is also presented
for comparison.  The intrinsic QD lifetime $\tau_{0}$ does not
vary between QDs in the unprocessed material and in the
unpatterned GaAs membrane.  It increases weakly from
$\tau_{0}\sim0.65 - 0.90ns$ as the detection wavelength increases
from $1000-1040nm$. From the data presented Fig. 3 the significant
lengthening (off-resonance) and shortening (on-resonance) of the
decay liftime discussed above can be clearly observed.  Moreover,
the lengthening is found to occur over a remarkably wide bandwidth
($\Delta\lambda\geq40nm$) and becomes more pronounced towards
longer wavelength as shown schematically by the dotted line on
Fig. 3 that acts as a guide to the eye. For the presently
investigated PCs, the middle of the TE-polarized photonic bandgap
lies at $\lambda_{mid}^{TE}\sim1100$ nm (see Fig. 1c) and there is
no overlapping gap for both $TE$ and $TM$ polarized waveguide
modes. The observed spectral dependence is attributed to a
progressive reduction of the local photon density of states as the
wavelength approaches $\lambda_{mid}^{TE}$, indicating that even a
\textit{partial} 2D photonic bandgap is sufficient to
significantly inhibit spontaneous emission.  We believe that this
is due to the predominantly \textit{heavy} hole character of the
QD ground state exciton transition\cite{Fry00} that gives rise to
$TE$ polarised emission.  Therefore, we suggest that tailoring of
only the $TE$-optical modes is sufficient to strongly modify the
spontaneous emission properties of self-assembled QDs in 2D-PC
nanocavities.

The minima in the spectral dependence of the
decay lifetime ($\tau(\lambda)$) close to $M1$ and $M2$ are
$\sim$4$\times$ broader than the cavity modes in the emission
spectrum.  Since the Lorenztian cavity modes should lead to a
similar spectral profile in $\tau(\lambda)$, this observation
indicates that the reduction of the decay time for zero detuning
is much larger than the measured $\sim150$ ps, limited by
our temporal resolution. In the weak coupling regime photon
reabsorption can be neglected and Fermi's golden rule provides the
spontaneous decay time relative to its value in a homogeneous
medium $\tau_{0}/\tau_{2}$.\cite{Gerard98}

\begin{equation}
    \frac{\tau_{0}}{\tau_{2}}=\frac{1}{3} F_{P}    \frac{\left|\vec{E}(\vec{r})\right|^{2}}{\left|\vec{E}_{max}\right|^{2}}
    \frac{\Delta\lambda_{cav}^{2}}{\Delta\lambda_{cav}^{2}+4(\lambda_{cav}-\lambda_{QD})^{2}} +    \alpha
\end{equation}

In equation 1 $\lambda_{QD}$ and $\lambda_{cav}$ are the QD and
cavity wavelength and $\Delta\lambda_{cav}$ is the linewidth of
the cavity mode measured from the PL spectrum.  Two different
decay channels are taken into account in Eqn. 1; the first term
describes the spontaneous emission of a dot located at $\vec{r}$
into the cavity mode with an local electric field
$\vec{E}(\vec{r})$ and a maximum amplitude $\vec{E}_{max}$,
whereas in the second term $\alpha$ describes a possible decay
channel due to emission into residual modes in the quasi-photonic
bandgap.  By fitting Eqn. 1 to the measured $\tau(\lambda)$ data
we extract the decay time on resonance, the best fit is compared
with the PL intensity in Fig. 3(inset) on a logarithmic scale. The
fitted spectral dependence of the decay time, now has exactly the
same lineshape as the PL intensity but with a much shorter decay
time $\tau_{2}=44\pm8ps$ on resonance. From the fit, we estimate a
maximum shortening of the decay time by a factor
$\tau_{0}/\tau_{2}=19\pm4$ for ideally located QDs on resonance,
corresponding to a maximum Purcell factor $F_{P}=56\pm10$.  This
value is in fairly good agreement with the maximum value of
$\sim100$ calculated from the measured $Q$ and $V_{eff}$, the
discrepancy probably arising from a displacement of the QDs probed
relative to the electric field antinode in the cavity.  For the
obtained values of $\tau_{1}=1.8ns$ and $\tau_{2}\sim50ps$ we
estimate a maximum single mode coupling efficiency for this system
of $\beta\sim97\%$.

In summary, we have investigated the influence of the modified
density of states in PC microcavities on the spontaneous emission
dynamics of self-assembled QDs.  A reduction of the spontaneous
emission lifetime up to $(5.6\pm0.3)\times$ was directly measured
for dots on resonance with the cavity modes and a lengthening
$>2\times$ off resonance over a bandwidth $\geq40$ nm.  From the
spectral dependence, a maximum enhancement of $\tau_{0}/\tau_{2}$
up to $19\pm4$ was deduced, corresponding to a drastic shorting of
the exciton lifetime to only $\sim50$ps.  This indicates that the
single mode coupling efficiency may become as large as
$\beta\sim97\%$.  Finally, since the numbers reported here are
ensemble averages, they may become even larger for an ideally
placed dots with a dipole moment aligned perfectly with the cavity
field.  Therefore, the present results provide significant promise
for realisation of \textit{efficient} single photon emitters based
on PC nano-cavities.

The authors gratefully acknowledge financial support by Deutsche
Forschungsgemeinschaft via $SFB-631$ and J. Zarbakhsh and K.
Hingerl from the University of Linz for theoretical input.


\begin{thebibliography}{99}
\bibitem{modes} The electric field profiles for the two modes are shown as an inset on Fig. 1c, from which we deduce an effective mode volume of $V_{eff}\sim1.5(\lambda/n)^{3}$.
\end{thebibliography}
\end{document}